\documentclass[aps,prd,twocolumn,nofootinbib,superscriptaddress,floatfix,notitlepage,10pt]{revtex4-1}

\usepackage{graphicx}
\usepackage{amsmath,amsfonts,amssymb}
\usepackage[dvipsnames]{xcolor}
\usepackage[breaklinks,colorlinks,urlcolor=Plum,citecolor=RoyalBlue,linkcolor=RubineRed]{hyperref}

\newcommand{\be}{\begin{equation}} 
\newcommand{\ee}{\end{equation}}
\newcommand{\bea}{\begin{equation}\begin{aligned}} 
\newcommand{\eea}{\end{aligned}\end{equation}}
\newcommand{\td}{{\rm d}}

\newcommand{\R}[1]{{\color{red} #1}}

\begin{document}

\title{Axion misalignment across first-order phase transitions}

\author{Galymzhan Baltabay}
\email{galymzhan.baltabay@kbfi.ee}
\affiliation{Laboratory of High Energy and Computational Physics, NICPB, R{\"a}vala 10, Tallinn, 10143, Estonia}
\affiliation{Department of Cybernetics, Tallinn University of Technology, Akadeemia tee 21, 12618 Tallinn, Estonia}

\author{Francesco D'Eramo}
\email{francesco.deramo@pd.infn.it}
\affiliation{Dipartimento di Fisica e Astronomia, Universit\`a degli Studi di Padova, Via Marzolo 8, 35131 Padova, Italy}
\affiliation{Istituto Nazionale di Fisica Nucleare, Sezione di Padova, Via Marzolo 8, 35131 Padova, Italy}

\author{Ville Vaskonen}
\email{ville.vaskonen@pd.infn.it}
\affiliation{Laboratory of High Energy and Computational Physics, NICPB, R{\"a}vala 10, Tallinn, 10143, Estonia}
\affiliation{Dipartimento di Fisica e Astronomia, Universit\`a degli Studi di Padova, Via Marzolo 8, 35131 Padova, Italy}
\affiliation{Istituto Nazionale di Fisica Nucleare, Sezione di Padova, Via Marzolo 8, 35131 Padova, Italy}

\begin{abstract}
When the axion mass is generated during a first-order phase transition and becomes nonvanishing only inside expanding true-vacuum bubbles, the standard picture of misalignment production is qualitatively modified. Using lattice simulations in an expanding universe, we study dark matter production within such a framework and identify two distinct  regimes. For rapid transitions, the onset of oscillations is delayed until bubble percolation, enhancing the relic abundance. For slower transitions, spatial gradients generated by expanding bubbles suppress the effective misalignment angle through the bubble misalignment mechanism. We derive a semianalytical expression for the relic density that provides a unified description of both regimes and accurately reproduces the simulation results. Finally, we show how this mechanism also modifies isocurvature perturbations and the small-scale matter power spectrum, with important implications for axion minicluster formation.
\end{abstract}

\maketitle

\section{Introduction}
\label{sec:intro}

New light and weakly coupled degrees of freedom arise naturally in well-motivated extensions of the Standard Model (SM) and, in the coming decades, will be probed with unprecedented sensitivity at both the intensity and cosmic frontiers. Among the candidates, pseudo-Nambu-Goldstone bosons (PNGBs) are particularly compelling, as their origin explains both their small masses and feeble interactions without requiring ad hoc assumptions. A prominent realization of this idea is provided by the Peccei-Quinn (PQ) mechanism~\cite{Peccei:1977hh,Peccei:1977ur}, originally proposed to solve the strong $CP$ problem. Following the spontaneous breaking of the PQ symmetry and the decoupling of the heavy PQ-sector states, the low-energy spectrum contains a PNGB known as the QCD axion~\cite{Wilczek:1977pj,Weinberg:1977ma}. Other well-motivated examples include the Majoron, associated with the spontaneous breaking of lepton number~\cite{Chikashige:1980ui}, as well as the rich spectrum of axionlike particles generically arising in string compactifications~\cite{Svrcek:2006yi,Arvanitaki:2009fg,Cicoli:2012sz}.

The cosmological evolution of the axion field is typically divided into two main phases. Initially, the field is displaced from the minimum of its potential. Its value may be set, for instance, by inflationary dynamics, while Hubble friction keeps the field frozen. At later times, the restoring force due to the potential eventually overcomes Hubble friction, causing the field to undergo damped oscillations around the minimum of the potential. Once the oscillation amplitude has been sufficiently damped, the dynamics can be described by harmonic oscillations. Crucially, the energy stored in these oscillations redshifts as nonrelativistic matter. Axions produced through this misalignment mechanism therefore provide an irreducible contribution to the present-day energy density~\cite{Preskill:1982cy,Abbott:1982af,Dine:1982ah}.

Evidence for a dark matter (DM) component with an abundance roughly five times larger than that of visible matter is overwhelming~\cite{Bertone:2004pz,Cirelli:2024ssz}. In this context, the question is not whether axions contribute to the DM energy density, but rather what fraction of the DM abundance they account for~\cite{Turner:1985si,Sikivie:2006ni,Marsh:2015xka}. The quantitative answer depends on both the particle physics parameters and the cosmological history. The case of the QCD axion with a standard expansion history is predictive because its mass and interactions are not independent parameters. In the deconfined QCD phase, the axion acquires a thermal mass with a power-law dependence on the temperature, $m_a \propto T^{-4}$. After confinement, the mass reaches its asymptotic value, $m_a \simeq \Lambda_{\rm QCD}^2/f_a$, where $\Lambda_{\rm QCD}$ and $f_a$ denote the confinement scale and the axion decay constant, respectively. The resulting relic density depends on the initial field value, which may be determined by the interplay with inflationary physics. If PQ symmetry breaking occurs after inflation, the standard prescription is to average over possible initial field values, yielding a misalignment contribution to the axion DM density that depends only on $f_a$. In this case, however, topological defects also contribute to the DM abundance~\cite{Vilenkin:1982ks,Kibble:1984hp,Harari:1987ht,Shellard:1987bv,Hagmann:1990mj,Battye:1994au,Kawasaki:2018bzv,Gorghetto:2018myk,Buschmann:2019icd,Hindmarsh:2019csc,Gorghetto:2020qws,Buschmann:2021sdq,Kim:2024wku,Kim:2024dtq,Benabou:2024msj}.

In this work, we explore the axion field dynamics in cosmological scenarios featuring abrupt departures from the standard expansion history. Specifically, we investigate the impact of first-order phase transitions (FOPTs) on the axion relic density. The key idea can be appreciated from the fact that the axion mass in our scenario turns on discontinuously inside expanding bubbles of true vacuum during a FOPT, unlike the smooth temperature-dependent evolution exhibited by the QCD axion. This leads to qualitatively different axion dynamics~\cite{Hindmarsh:1991ay,Nakagawa:2022wwm,Lee:2024oaz,Carenza:2024tmi,GarciaGarcia:2024dfx,Bhandari:2025pzg,Bhandari:2025phe}.

We focus on a general framework in which an axion field $\phi$ has a mass that vanishes in the false vacuum while remaining finite in the true vacuum. The dynamics is trivial before the FOPT: the field remains frozen since no potential is present. After the FOPT, the axion mass is switched on and the field can undergo damped oscillations. The behavior for sufficiently fast FOPTs is intuitive. In this case, the onset of the oscillatory phase is delayed, leaving less time for the initial potential energy to redshift and thereby enhancing the DM relic abundance~\cite{Nakagawa:2022wwm}. For slower transitions, however, spatial gradients generated by the expanding bubble walls suppress the oscillation amplitude inside the bubbles. This regime was investigated in Ref.~\cite{Lee:2024oaz}, where the resulting mechanism was dubbed bubble misalignment. In particular, the authors simulated the evolution of an axion field with a discontinuous harmonic potential in a simplified cosmological setup, neglecting Hubble expansion and considering a single bubble nucleated at the center of the simulation box with periodic boundary conditions.

We investigate the interplay between axion dynamics and FOPTs in a more realistic cosmological setting. In particular, we incorporate the effects of cosmic expansion, allowing us to consistently follow the redshifting of both field gradients and oscillation amplitudes. We also account for the  nucleation and evolution of multiple bubbles via a self-consistent nucleation history. Anharmonic effects in the axion potential, important in the standard QCD axion scenario~\cite{Preskill:1982cy,Turner:1985si,Lyth:1991ub,Bae:2008ue,Visinelli:2009zm}, are likewise included. Together, these ingredients provide a substantially more realistic characterization of the resulting dynamics.

FOPTs affect axion dynamics in several observable ways. The delayed onset of oscillations modifies the dependence of the relic abundance on the axion mass and decay constant. In the postinflationary scenario, the phase transition lowers the effective cutoff scale of the white-noise isocurvature spectrum, enhancing the mass and abundance of axion miniclusters. In the preinflationary scenario, it modifies the mapping between inflationary axion fluctuations and the relic abundance, leading to modified CMB isocurvature bounds in both stochastic and locked initial-condition regimes. These effects are constrained by Lyman-$\alpha$ forest data and JWST observations of the high-redshift UV luminosity function.

Our quantitative results apply directly to generic axionlike particles whose mass is generated through a new FOPT. The minimal QCD axion does not fall directly within this framework: the thermal QCD transition is a crossover, and the topological susceptibility generates a smooth, nonzero axion mass already in the deconfined phase, rather than an abrupt mass generation during a FOPT. Extending the present analysis to the QCD axion would therefore require an additional FOPT coupled to the QCD sector, whose contribution to the axion mass dominates over the standard topological susceptibility in the relevant temperature range. We leave this extension for future work.

Section~\ref{sec:bubble} presents a description of true-vacuum bubble nucleation. The axion setup is introduced in Sec.~\ref{sec:axion}, where we derive the field equations governing the dynamics. We employ a low-energy scalar potential including anharmonic effects that captures a broad class of UV-complete realizations.  Appendix~\ref{app:model} discusses an explicit class of renormalizable UV-complete models that naturally give rise to the low-energy phenomenology explored in this work. Section~\ref{sec:abundance} characterizes the two dynamical regimes, constructs a semianalytic model for the relic abundance. We also show that axions radiated by cosmic strings are suppressed relative to the misalignment contribution when the mass turns on abruptly. The lattice implementation is validated against homogeneous solutions in Appendix~\ref{app:validation}. In Sec.~\ref{sec:pheno}, we quantify the observational signatures and derive constraints from axion minicluster formation, CMB isocurvature perturbations, Lyman-$\alpha$ forest data, JWST observations of high-redshift galaxies, and black-hole superradiance. We summarize our findings in Sec.~\ref{sec:conclusion}.

\section{Bubble Dynamics}
\label{sec:bubble}

The cosmological evolution in our setup begins with the universe trapped in a metastable false-vacuum state. As the universe expands and cools, bubbles of true vacuum nucleate and grow within the false-vacuum background, eventually percolating and completing the FOPT. Once the microscopic details of the sector responsible for the FOPT are specified, the bubble nucleation rate can be computed from the finite-temperature effective potential~\cite{Coleman:1977py,Callan:1977pt,Linde:1981zj}.

We remain agnostic about the underlying microphysics responsible for the FOPT and adopt a phenomenological approach. Specifically, we introduce the reference time $t_\star$ as the epoch at which the nucleation rate reaches one bubble per Hubble volume and Hubble time.\footnote{Throughout this paper, the variable $t$ denotes the time coordinate of the Friedmann-Lema\^{\i}tre-Robertson-Walker (FLRW) metric. We will later also introduce conformal time $\tau$.} Denoting by $H_\star$ the Hubble parameter evaluated at this time, $H_\star \equiv H(t_\star)$, we parametrize the nucleation rate per unit volume through the exponential law~\cite{Turner:1992tz}
\be \label{eq:Gamma}
    \Gamma(t) = H_\star^4 e^{\beta (t-t_\star)} \ ,
\ee
where the parameter $\beta$ characterizes the inverse duration of the phase transition, with large values of $\beta$ corresponding to rapid nucleation. This expression can be understood as the linear expansion of the logarithm of the nucleation rate around the transition epoch, and it often provides a good approximation to the time dependence of the nucleation rate in realistic models (see, e.g.,~\cite{Lewicki:2024sfw}). By construction, Eq.~\eqref{eq:Gamma} describes a transition that effectively begins at $t=t_\star$, when the nucleation rate reaches one bubble per Hubble volume and Hubble time.\footnote{From the definition in Eq.~\eqref{eq:Gamma}, we have $\Gamma(t_\star)=H_\star^4$.}

Throughout this work, we assume the early universe to be dominated by a radiation bath of relativistic particles, for which the scale factor evolves as $a(t) \propto t^{1/2}$. We normalize the scale factor to unity at the time of the transition, namely $a_\star \equiv a(t_\star)=1$, and choose $t_\star = 1/(2 H_\star)$. This allows us to express the scale factor as $a(t) = (2 H_\star t)^{1/2}$ and the Hubble rate as $H(t) = 1/(2 t)$. For the cosmological simulations, it is more convenient to work with conformal time $\tau$, related to cosmic time through $d\tau = dt/a(t)$, which in a radiation-dominated background implies $\tau(t) = (2 t/H_\star)^{1/2}$. In terms of conformal time, the scale factor takes the simple form $a(\tau) = H_\star \tau$ and the Hubble rate is $H(\tau) = 1/(H_\star \tau^2)$.

Once nucleated, bubbles expand within the surrounding false vacuum background, driven by the pressure difference across the wall. In our analysis, we neglect their initial radii and work in the thin-wall limit, in which the bubble walls are treated as infinitesimally thin interfaces separating the two phases. In the absence of interactions with the surrounding fluid, the bubble walls rapidly approach the speed of light, $v_w \approx 1$, while their Lorentz factors grow linearly with the bubble radius. Achieving a smaller terminal velocity requires frictional effects arising from interactions between the walls and the surrounding plasma. If sufficiently strong, these effects balance the pressure difference driving the expansion, causing the walls to approach a steady-state configuration characterized by a subluminal terminal velocity~\cite{Ignatius:1993qn,Moore:1995si,Bodeker:2017cim,Krajewski:2024xuz,Ekstedt:2024fyq}.

\begin{figure}
    \centering
    \includegraphics[width=1.0\linewidth]{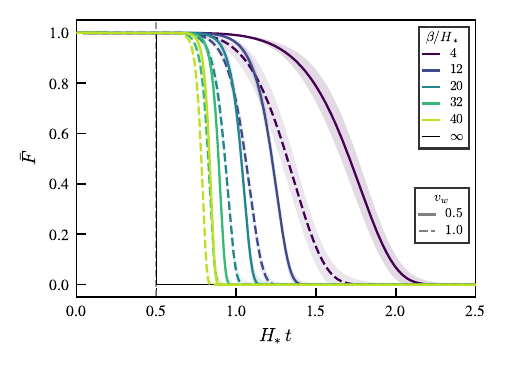}
    \caption{Evolution of the false-vacuum fraction $\bar{F}(t)$ for different values of $\beta/H_\star$ and $v_w$ in a radiation-dominated universe. The solid curves correspond to the numerical evaluation of Eq.~\eqref{eq:barF}, while the shaded bands indicate the $1\sigma$ spread among different simulation realizations.}
    \label{fig:fvf_vs_time}
\end{figure}

Following our phenomenological approach, we do not commit to any specific microscopic origin for the friction effects and instead treat the bubble wall velocity as a free parameter. In particular, we assume that the bubbles expand with a constant wall velocity $v_w \leq 1$ throughout the transition. In terms of the conformal time variable $\tau$ introduced above, the comoving bubble radius therefore evolves linearly according to~\footnote{Notice that $v_w$ is defined as the bubble wall velocity with respect to the comoving bubble radius and conformal time, $v_w \equiv \td R/\td \tau$.}
\be \label{eq:bubbleradius}
    R(\tau;\tau_n) = \max[0,v_w(\tau-\tau_n)] \,, 
\ee
where $\tau_n$ denotes the conformal nucleation time.\footnote{The time $t_\star$ appearing in Eq.~\eqref{eq:Gamma} and the nucleation time $t_n$ are distinct quantities. The former is fixed once a given FOPT is specified, whereas the latter is a variable corresponding to the nucleation time of the particular bubble under consideration.} The nucleation and subsequent expansion of bubbles progressively reduce the fraction $\bar{F}(t)$ of the universe that remains trapped in the false vacuum, which is given by~\cite{Guth:1982pn}
\be \label{eq:barF}
    \bar{F}(t) = \exp\left[-\frac{4\pi}{3} \!\int_{-\infty}^t \!\td t_n \Gamma(t_n) a(t_n)^3 R(t;t_n)^3\right] \,.
\ee
The integral over $t_n$ accounts for all nucleation times. The integrand contains the nucleation rate $\Gamma(t_n)$ together with the physical volume of a bubble nucleated at time $t_n$, whose radius is determined by Eqs.~\eqref{eq:Gamma} and \eqref{eq:bubbleradius}, respectively. We show the resulting evolution of $\bar{F}(t)$ in Fig.~\ref{fig:fvf_vs_time} for different values of $\beta/H_\star$ and $v_w$. As expected, larger values of $\beta/H_\star$ and $v_w$ correspond to increasingly rapid transitions. In the limit $\beta/H_\star \to \infty$, the transition becomes instantaneous and the false-vacuum fraction approaches the step-function behavior $\bar{F}(t)\to \Theta(t_\star-t)$.

We perform simulations of the FOPT by nucleating bubbles within a cubic box of volume $L_{\rm box}^3$ with periodic boundary conditions. Bubble nucleation follows the rate~\eqref{eq:Gamma} and is implemented using a tracer-based method. A total of $N_{\rm tracer} = 10^5$ tracer points are distributed uniformly throughout the box, each representing a potential nucleation site in the false vacuum. At each time nucleation step of duration $\Delta t$, a random number $P \sim U(0,1)$ is drawn for each tracer still residing in the false vacuum, and a bubble is nucleated at that tracer's position if $P < P_{\rm nuc}(t) = \Gamma(t) a(t)^3 L_{\rm box}^3 \Delta t / N_{\rm tracer}$. The nucleation time step is chosen adaptively to ensure that $P_{\rm nuc} < 0.1$. Once a tracer nucleates a bubble or is engulfed by an expanding bubble, it is removed from the active set. The false-vacuum fraction $\bar{F}(t)$ is then estimated as the ratio of active to total tracers. Figure~\ref{fig:fvf_vs_time} compares the simulation results with the semianalytic prediction of Eq.~\eqref{eq:barF}. The agreement between the two approaches is found to be excellent across a wide range of parameters.
We note that for transitions that are too fast or for box sizes that are too large, the tracer-based method may effectively run out of active tracer points before the transition completes. This imposes an upper limit on both $\beta/H_\star$ and $L_{\rm box}$. We have verified that $N_{\rm tracer}=10^5$ is sufficient for the parameter range considered in this work. The validity range can be extended by increasing $N_{\rm tracer}$, at the expense of additional computational cost.

\section{Axion Dynamics}
\label{sec:axion}

The dynamics of the axion field are governed by the periodic scalar potential
\be \label{eq:axion_potential}
    V(\phi) = m_\phi(S)^2 f_\phi^2 \left[ 1-\cos\left(\frac{\phi}{f_\phi}\right) \right] \,.
\ee
Here, $f_\phi$ denotes the axion decay constant and is associated with the scale of spontaneous breaking of the underlying global symmetry. We take the axion field to have support $\phi \in [0,2\pi f_\phi)$, such that the potential possesses a single physically distinct minimum at $\phi=0$.\footnote{The case $\phi \in [0,2\pi N_{\rm DW}f_\phi)$ with integer $N_{\rm DW}>1$ is more subtle, as additional bias terms are generally required to lift the vacuum degeneracy and prevent cosmological domain-wall domination. For this reason, we restrict our analysis to $N_{\rm DW}=1$.} Expanding Eq.~\eqref{eq:axion_potential} around its minimum to quadratic order identifies $m_\phi(S)$ as the axion mass. As indicated by the notation, we allow this mass to depend on a scalar field $S$. Explicit renormalizable realizations that reproduce the potential in Eq.~\eqref{eq:axion_potential} at low energies are presented in Appendix~\ref{app:model}.

The field $S$ that determines the axion mass is also the degree of freedom responsible for driving the FOPT. We focus on scenarios in which the axion mass vanishes before the transition. Assuming the false vacuum is located at $S=0$, this condition implies $m_\phi(S=0)=0$. In the true vacuum the scalar field $S$ acquires a nonvanishing vacuum expectation value (vev), $\langle S \rangle = v_S$, and we denote the resulting axion mass by $M_\phi$. In the thin-wall limit, the dynamics are determined solely by the values of the axion mass in the two phases:
\be \label{eq:axionmassS}
    m_\phi(S) =
    \begin{cases}
        0 \,, & S=0 \\
        M_\phi \,, & S=v_S
    \end{cases} \,.
\ee

Before the phase transition, the axion field $\phi$ within a given Hubble patch is frozen at a random initial value. In our analysis, we allow for arbitrary initial conditions in the range $0 \leq \phi(t=0) < 2\pi f_\phi$, which may vary from one Hubble patch to another.

The dynamics of the axion field $\phi$ are governed by the covariant equation of motion
\be
    \nabla_\mu \nabla^\mu \phi \equiv \frac{1}{\sqrt{-g}} \partial_\mu \left( \sqrt{-g} \, g^{\mu\nu}\partial_\nu \phi \right) = \frac{\partial V(\phi)}{\partial \phi} \,.
\ee
Expressing the field as $\phi=\phi(t,\vec{x})$, where $t$ denotes cosmic time and $\vec{x}$ comoving spatial coordinates, and specializing to the FLRW metric and the potential in Eq.~\eqref{eq:axion_potential}, the equation of motion takes the form
\be
    \partial_t^2 \phi + 3H\,\partial_t \phi - \frac{\nabla^2\phi}{a^2} + f_\phi m_\phi^2(S)\sin\left(\frac{\phi}{f_\phi}\right) = 0 \,.
\ee

It is convenient to reformulate the axion dynamics in terms of the dimensionless field $\theta \equiv \phi/f_\phi$ and conformal time $\tau$. The equation of motion becomes
\be \label{eq:eoftau}
    \partial_\tau^2 \theta + 2\mathcal{H} \partial_\tau\theta - \nabla^2\theta + a^2 m_\phi^2(S)\sin\theta = 0 \,,
\ee
where $\mathcal{H} \equiv (\partial_\tau a)/a$ is the conformal Hubble parameter, related to the standard one through $\mathcal{H}=aH$. For the radiation-dominated background considered in this work, and with the conventions of Sec.~\ref{sec:bubble}, the conformal Hubble parameter is given by $\mathcal{H}=1/\tau$. Using the field dependence of the axion mass given in Eq.~\eqref{eq:axionmassS} and assuming that $S$ vanishes outside the expanding bubbles, the equation of motion reads
\be \label{eq:eom_conformal}
    \partial_\tau^2 \theta + \frac{2}{\tau}\partial_\tau \theta - \nabla^2 \theta + H_\star^2\tau^2 M_\phi^2 \frac{\partial \mathcal{V}(\theta,\vec{x},\tau)}{\partial \theta} = 0 \,.
\ee
In the thin-wall approximation, the dimensionless potential is
\be \label{eq:Vdimless} 
    \mathcal{V}(\theta,\vec{x},\tau) = 
    \begin{cases} 
        1 - \cos\theta \,, & \min_j \dfrac{|\vec{x} - \vec{x}_j|}{R_j(\tau)} < 1 \,, \\[4pt] 
        0 \,, & \text{otherwise}\, , 
    \end{cases} 
\ee
with the index $j$ running over all nucleated bubbles, while $\vec{x}_j$ and $R_j(\tau)$ denote their centers and radii, respectively.

We evolve the axion field on a periodic lattice with $N_{\rm grid}^3$ grid points covering the same cubic volume used for bubble nucleation. To ensure compatibility between the periodic boundary conditions and the axion oscillations, we choose the box size $L_{\rm box}$ to be an integer multiple of the axion Compton wavelength $\lambda_\phi = 2\pi/M_\phi$. The simulation volume is initially in the false vacuum, while the axion field is uniformly initialized to the value $\theta_i$. The field is then evolved according to Eq.~\eqref{eq:eom_conformal} using a leapfrog algorithm~\cite{Figueroa:2021yhd}, with spatial derivatives computed through real-space finite-difference stencils. We do not employ pseudospectral or Fourier-based (FFT) solvers, since the discontinuous step-function profile of the bubble walls in Eq.~\eqref{eq:Vdimless} gives rise to a severe Gibbs phenomenon. This generates spurious ringing artifacts that contaminate the field evolution and energy densities with unphysical high-frequency modes. The integration time step is determined by the Courant-Friedrichs-Lewy stability condition
\be \label{eq:CFL}
    \delta t = 0.4\left[ 3\left(\frac{\pi}{\Delta x}\right)^2 + M_\phi^2 \right]^{-1/2},
\ee
where $\Delta x=L_{\rm box}/N_{\rm grid}$ denotes the lattice spacing.

One of the key observables in our analysis is the present-day axion relic density. To determine it, we track throughout the simulation the local axion energy density
\be
    \rho_\phi(\vec{x}, t) = \frac{1}{2} (\partial_t \phi)^2 + \frac{1}{2 a^2} (\nabla \phi)^2 + V(\phi) \, .
\ee
Our first objective is to determine whether the observed DM abundance can be reproduced. We therefore define the spatial average of the previous expression as $\bar{\rho}_\phi(t) \equiv \langle \rho_\phi(\vec{x}, t) \rangle$, where $\langle \cdot \rangle$ denotes spatial averaging. Expressed in terms of conformal time $\tau$ and the dimensionless field $\theta$, this becomes
\be \label{eq:averaged-energy-density}
    \bar{\rho}_\phi = \frac{f_\phi^2}{H_\star^2} \left[ \frac{\langle(\partial_\tau \theta)^2\rangle + \langle(\nabla \theta)^2\rangle}{2 \tau^2} + H_\star^2 M_\phi^2 \langle\mathcal{V}(\theta)\rangle \right] \,, 
\ee
where $\mathcal{V}(\theta,\vec{x},\tau)$ is the potential introduced in Eq.~\eqref{eq:Vdimless}.

We validated our numerical implementation through homogeneous-field tests in an expanding radiation-dominated background. We initialized the axion field to a spatially uniform configuration across the lattice and assumed a constant nonzero mass $M_\phi$, such that spatial gradients vanish identically throughout the evolution. In the small-amplitude regime, where the scalar potential is well approximated by its harmonic expansion, the equation of motion admits an exact analytic solution in terms of Bessel functions. We verified that the lattice evolution reproduces this solution with subpercent accuracy over several oscillation periods. In the nonlinear regime, we further confirmed conservation of the late-time adiabatic invariant $\bar{\rho}_\phi a^3$, recovering the expected matterlike scaling once the condition $M_\phi \gg H$ is satisfied. Details of these validation tests are presented in Appendix~\ref{app:validation}.

Figure~\ref{fig:energy_evolution} shows the results of simulations performed with $L_{\rm box}=8\pi/M_\phi$ and $N_{\rm grid}=128$. We fix $\beta/H_\star=4$, $H_\star/M_\phi=0.05$, $v_w=0.15$, and $\theta_i=2.0$. The figure shows the evolution of the spatially averaged axion energy densities, with both axes rendered dimensionless through appropriate rescalings. The kinetic, gradient, potential, and total contributions are displayed separately. For the latter, we compare the evolution in the presence of the FOPT with that in a standard radiation-dominated background without a FOPT.

\begin{figure}
    \centering
    \includegraphics[width=\linewidth]{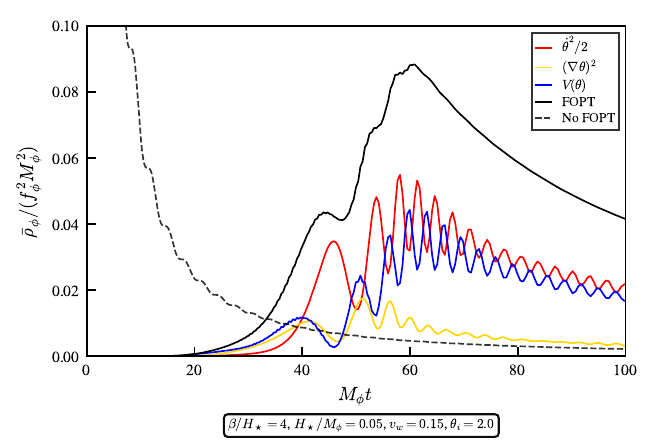}
    \caption{Spatially averaged axion energy density as a function of cosmic time. Both axes are expressed in dimensionless units after appropriate rescalings. The solid curves show the evolution in the presence of a FOPT, with the kinetic (red), gradient (yellow), and potential (blue) energy densities displayed separately, while the total energy density is shown in black. For comparison, the dashed curve shows the total axion energy density in the absence of a FOPT.}
    \label{fig:energy_evolution}
\end{figure}

Before the phase transition, the axion potential vanishes and the field remains frozen at its initial value. During the FOPT, bubbles nucleate at an exponentially increasing rate, and the potential energy rises rapidly as an increasing fraction of the volume transitions to the true vacuum, where the axion acquires the mass $M_\phi$. After the transition, the gradient energy redshifts as $a^{-4}$, while the combined kinetic and potential contributions scale as matter, namely $\propto a^{-3}$. The latter agrees with the standard evolution in a radiation-dominated universe without a phase transition, shown by the dashed curve. The most interesting dynamics occur during the phase transition itself, when the energy density exhibits oscillatory growth due to repeated reflections of axion waves from the expanding bubble walls. This behavior resembles that found in Ref.~\cite{Lee:2024oaz} for a static background, and it will be further investigated in Sec.~\ref{subsec:bubble_m}.

\section{Axion relic density}
\label{sec:abundance}

The first observable quantity we consider is the axion relic density. Its importance is already apparent from Fig.~\ref{fig:energy_evolution}, particularly from the comparison of the two black curves showing the evolution of the spatially averaged total axion energy density with and without a FOPT (solid and dashed, respectively). As discussed in the previous section, their early time evolution differs significantly due to the nontrivial dynamics induced during the phase transition. At sufficiently late times, however, both curves asymptotically approach the matterlike scaling, $\bar{\rho}_a \propto a^{-3}$, implying that the comoving axion energy density becomes conserved. This allows for a direct comparison of the final relic abundance in the two scenarios.

A viable DM candidate must reproduce the observed relic density, $\Omega_{\rm DM} h^2 = 0.120 \pm 0.001$~\cite{Planck:2018vyg}, if it accounts for all of the DM. Smaller relic densities are also allowed, corresponding to a subdominant DM component, whereas larger values are excluded.

\subsection{Misalignment without a FOPT}
\label{subsec:standmis}

Before discussing the impact of a FOPT, it is useful to briefly review the conventional misalignment mechanism. This serves to introduce the notation and conventions used throughout the paper and to validate our numerical simulations by demonstrating that they correctly reproduce the well-established results in the literature. In this scenario, the axion field is initially frozen and starts oscillating once the Hubble damping becomes comparable to the restoring force. Neglecting anharmonic corrections, the onset of oscillations is defined, up to factors of order unity, by $H(t_{\rm osc}) \sim M_\phi$, corresponding to $t_{\rm osc} \sim M_\phi^{-1}$ during radiation domination.

The axion energy density at $t_{\rm osc}$ is given by $M_\phi^2 f_\phi^2 \theta_i^2/2$, where $\theta_i$ denotes the initial misalignment angle and we still work within the harmonic approximation. The present-day relic abundance is obtained by evolving the energy density at the onset of oscillations to late times using conservation of the adiabatic invariant, which suppresses it by a factor $[a(t_{\rm osc})/a(t_0)]^3$, where $t_0$ denotes the present epoch. Using entropy conservation, we obtain $\bar{\rho}^{\rm NoFOPT}_{\phi,\mathrm{harmonic}}(t_0) \propto \theta_i^2 f_\phi^2 M_\phi^{1/2}$.

We evaluate the fractional spatially averaged energy density $\Omega_{\phi} \equiv \bar{\rho}_\phi / \rho_{\rm cr}$ directly from the output of our numerical simulations, where the present-day critical density is $\rho_c(t_0) \simeq 1.054 \times 10^{-5} h^2~{\rm GeV}/{\rm cm}^3$~\cite{ParticleDataGroup:2024cfk}. We find
\bea \label{eq:Omega_noPT}
    \Omega_\phi^{\rm NoFOPT} h^2 \simeq & \, 1.0 \, \frac{g_\star(T_{\rm osc})^{3/4}}{g_{*s}(T_{\rm osc})} \mathcal{F}_{\rm anh}(\theta_i) \theta_i^2 \\ 
    & \qquad \quad \times \left[\frac{f_\phi}{10^{12}{\rm GeV}}\right]^2 \left[\frac{M_\phi}{\rm eV} \right]^{1/2} ,
\eea
where the overall normalization is fixed by our numerical simulations. Here, $g_\star(T)$ and $g_{*s}(T)$ denote the effective numbers of relativistic degrees of freedom contributing to the energy and entropy densities, respectively.

Following the standard convention, we account for anharmonic corrections by multiplying the harmonic result by a correction factor $\mathcal{F}_{\rm anh}(\theta_i)$. As is well known, anharmonic effects enhance the axion energy density relative to the harmonic prediction~\cite{Preskill:1982cy,Turner:1985si,Lyth:1991ub,Bae:2008ue,Visinelli:2009zm}. Physically, this enhancement arises because the restoring force becomes significantly weaker near the top of the potential, delaying the onset of oscillations. The axion field therefore remains frozen for longer and undergoes less cosmological dilution before behaving as cold matter. Consequently, $\mathcal{F}_{\rm anh}(\theta_i)$ must reproduce the harmonic result in the small-angle limit, $\mathcal{F}_{\rm anh}(\theta_i \ll 1) \simeq 1$, while diverging as $\theta_i \rightarrow \pi$, reflecting the substantial delay in the onset of oscillations for fields initially located near the top of the potential. Fitting our numerical results, we obtain
\be \label{eq:fanh}
    \mathcal{F}_{\rm anh}(\theta_i)
    = \left[ \ln \frac{e}{1 - \left(|\theta_i|/\pi\right)^{\epsilon_0} } \right]^{\alpha_0} ,
\ee
with best-fit values $\epsilon_0 = 2.72$ and $\alpha_0 = 1.27$, obtained from simulations spanning the range $0 \leq \theta_i < \pi$.\footnote{We adopt the functional form proposed in Ref.~\cite{Visinelli:2009zm}. Our best-fit values differ from those reported there because that work considered the QCD axion with its temperature-dependent mass.}

We now discuss the impact of a FOPT. Before doing so, however, it is useful to identify the region of parameter space where the phase transition can affect the axion dynamics. If the axion acquires its mass through a FOPT with $H_\star \gg M_\phi$, Hubble friction still dominates at the time of the transition. Consequently, the axion remains frozen until well after the phase transition has completed, and its relic abundance coincides with the standard result in Eq.~\eqref{eq:Omega_noPT}. We therefore focus on the regime $H_\star \lesssim M_\phi$, where the phase transition occurs sufficiently late to modify the axion dynamics. The remaining parameter controlling the dynamics is $\beta$, introduced in the nucleation rate of Eq.~\eqref{eq:Gamma}. As discussed previously, $\beta^{-1}$ sets the characteristic duration of the phase transition. The hierarchy between this timescale and the axion oscillation period, $M_\phi^{-1}$, gives rise to two qualitatively distinct cosmological scenarios, discussed in the following two subsections.

\subsection{Fast FOPT ($\beta \gg M_\phi$): Delayed misalignment}
\label{subsec:delay}

We begin by discussing the case in which the phase transition completes on a timescale much shorter than the axion oscillation period. In this regime, characterized by the hierarchy $\beta \gg M_\phi$, the true-vacuum bubbles collide before the axion begins to oscillate inside them, and the field therefore does not respond dynamically during the transition. The dominant effect is instead a delay in the onset of axion oscillations, which enhances the present-day relic abundance.

A convenient reference time for the onset of oscillations in this case is the percolation time $t_p$, defined by
\be \label{eq:tp}
    \bar{F}(t_p) = e^{-1} \,,
\ee
where $\bar{F}(t)$ denotes the volume-averaged false-vacuum fraction. Since percolation occurs after the conventional oscillation time, $t_p \gtrsim t_{\rm osc} \sim M_\phi^{-1}$, the onset of oscillations is well approximated by $t_p$ rather than $t_{\rm osc}$.

If the phase transition is very fast, $\beta/H_\star\gg1$, percolation occurs shortly after the nucleation time, $t_p \simeq t_\star$, and the resulting enhancement becomes independent of $\beta/H_\star$. Conversely, for slower transitions, $\beta/H_\star\sim1$, the percolation time is significantly delayed, as illustrated in Fig.~\ref{fig:fvf_vs_time}. The corresponding enhancement of the present-day relic abundance is approximately given by the factor $(a_p/a_{\rm osc})^3 \sim (M_\phi/H_p)^{3/2}$, with $a_p \equiv a(t_p)$ and $a_{\rm osc} \equiv a(t_{\rm osc})$. In deriving the second expression, we assumed a radiation-dominated universe, where $H\propto a^{-2}$, used the relation $H(t_{\rm osc})\simeq M_\phi$, and defined $H_p\equiv H(t_p)$.

The estimate above, however, does not fully capture the enhancement associated with delayed misalignment. It assumes that the only effect of generating the axion mass through a FOPT is to delay the onset of oscillations. At later times, however, Hubble friction is weaker relative to the axion potential, so the anharmonic correction is no longer universal. Instead, it acquires an explicit dependence on the percolation time $t_p$.

\begin{figure}
    \centering
    \includegraphics[width=1.0\linewidth]{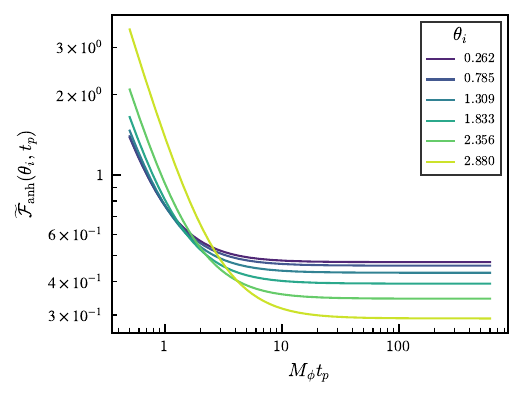}
    \caption{Anharmonic correction factor for delayed misalignment, $\widetilde{\mathcal{F}}_{\rm anh}(\theta_i,t_p)$, as a function of the percolation time for different initial axion misalignment angles.}
    \label{fig:f-tilde}
\end{figure}

To quantify this effect, we solve the homogeneous equation of motion with the axion mass switched on abruptly at $t=t_p$. The resulting anharmonic correction factor $\widetilde{\mathcal{F}}_{\rm anh}(\theta_i,t_p)$ is shown in Fig.~\ref{fig:f-tilde} as a function of the percolation time $t_p$. For $M_\phi t_p\ll1$, this result reproduces the standard anharmonic correction, $(M_\phi/H_p)^{3/2}\,\widetilde{\mathcal{F}}_{\rm anh}(\theta_i,t_p)\simeq\mathcal{F}_{\rm anh}(\theta_i)$, as expected in the conventional misalignment scenario without a FOPT. In the opposite limit, $M_\phi t_p\gg1$, Hubble friction is negligible at the onset of oscillations and the dynamics approaches that of an undamped field. The comoving energy is then conserved from the moment of release and is fixed by the initial potential energy, $\propto 1-\cos\theta_i$. Consequently $\widetilde{\mathcal{F}}_{\rm anh}$ approaches a finite, $t_p$-independent profile that \emph{decreases} with increasing $\theta_i$, since the cosine potential stores less energy than its quadratic approximation at large field values.

We find that the numerical results shown in Fig.~\ref{fig:f-tilde} are well described by the broken power law
\be \label{eq:tildefanh}
    \widetilde{\mathcal{F}}_{\rm anh}(\theta_i,t_p) = \left[C\,\mathcal{F}_\infty(\theta_i)^r + \frac{\mathcal{F}_{\rm anh}(\theta_i)^r} {\left(M_\phi/H_p\right)^{3r/2}}\right]^{1/r} \,,
\ee
where $\mathcal{F}_\infty(\theta_i)$ denotes the anharmonic correction factor in the limit of very late percolation. We parametrize it with the same functional form as the standard correction $\mathcal{F}_{\rm anh}$ in Eq.~\eqref{eq:fanh},
\be \label{eq:finfty}
    \mathcal{F}_\infty(\theta_i) = \left[\ln \frac{e}{1-\left(|\theta_i|/\pi\right)^{\epsilon_\infty}} \right]^{\alpha_\infty},
\ee
with the best-fit values $\epsilon_\infty=1.74$ and $\alpha_\infty=-0.46$, with $C=0.48$ and $r=1.12$ from the homogeneous evolution. In the presence of bubble nucleation, however, these parameters acquire a dependence on the bubble wall velocity, which we quantify in Sec.~\ref{subsec:fit}.

\subsection{Slow FOPT ($\beta \ll M_\phi$): Bubble misalignment}
\label{subsec:bubble_m}

\begin{figure}
    \centering
    \includegraphics[width=1.0\linewidth]{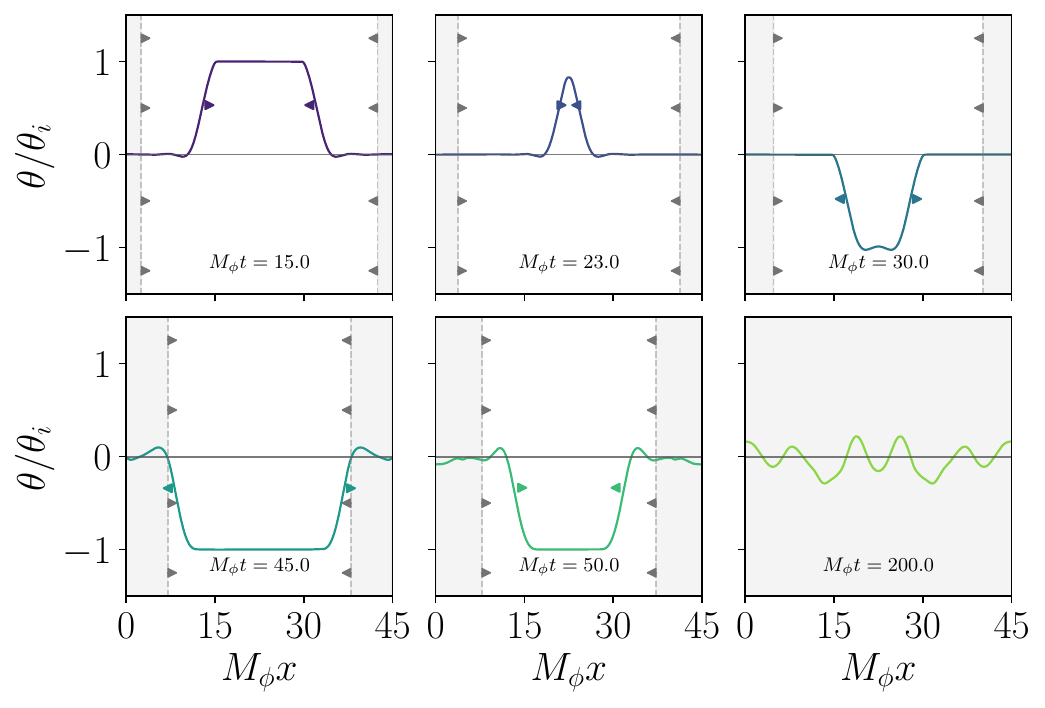}
    \caption{Evolution of the axion shock-wave profile for $\theta_i=2.0$. Each panel shows a snapshot at a different time, progressing from the top left to the bottom right. True-vacuum bubbles are nucleated simultaneously at $t=0$ at both ends of the simulation box, with the true-vacuum regions shaded in gray. The colored curves show the axion field profile, illustrating the formation and propagation of shock waves ahead of the expanding bubble walls.}
    \label{fig:shock}
\end{figure}

We now consider the complementary regime in which the phase transition is slow compared to the axion oscillation timescale, $\beta \ll M_\phi$. In this case, the transition is not complete before the axion begins to respond dynamically, leading to qualitatively different behavior. In particular, axion shock-wave profiles develop outside the expanding bubbles, as shown in Fig.~\ref{fig:shock}. Consequently, the axion field inside the bubbles begins oscillating with an amplitude smaller than its initial misalignment value, suppressing the relic abundance relative to the delayed-misalignment scenario. At the same time, repeated collisions between the shock waves and the expanding bubble walls transfer energy into the bubble interior, giving rise to the oscillatory growth of the axion energy density visible in Fig.~\ref{fig:energy_evolution}. This phenomenon is known as bubble misalignment~\cite{Lee:2024oaz}.

\begin{figure*}
\centering
\includegraphics[width=1\linewidth]{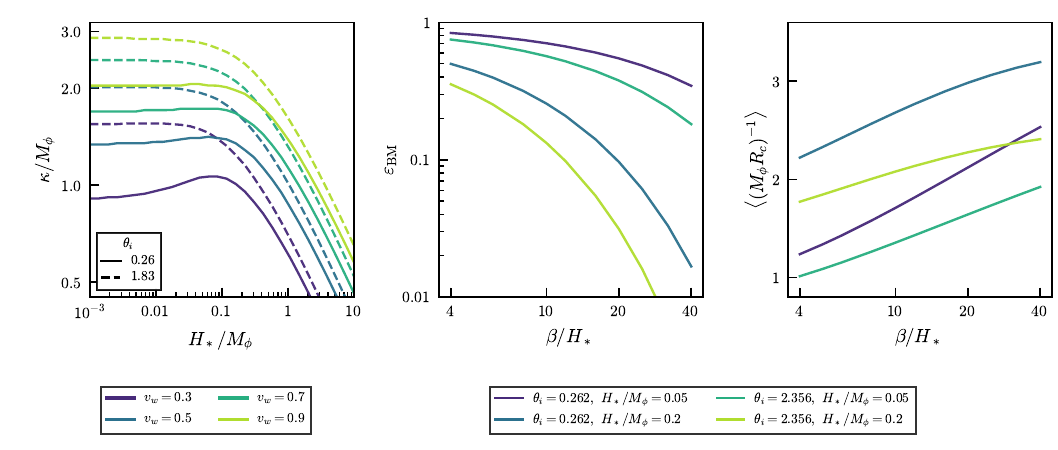}
\caption{Geometric quantities entering the semianalytic model for bubble misalignment. Left: minimum collision radius parameter $\kappa$ as a function of $H_\star/M_\phi$ for different bubble wall velocities $v_w$ and initial misalignment angles $\theta_i$. Middle: fraction of bubbles contributing to bubble misalignment, $\varepsilon_{\rm BM}$, as a function of $\beta/H_\star$. Right: mean inverse collision radius $\langle (M_\phi R_c)^{-1} \rangle$ as a function of $\beta/H_\star$.}
\label{fig:geom_plots}
\end{figure*}

Shock-wave formation is not guaranteed a priori and requires bubbles to reach a minimum size before colliding. We parametrize this requirement by introducing $\kappa$, such that only bubbles with collision radius $R_c>\kappa/M_\phi$ contribute to bubble misalignment. To determine the dependence of $\kappa$ on $v_w$ and $H_\star/M_\phi$, we solve Eq.~\eqref{eq:eom_conformal} for the evolution of the axion field around a single expanding bubble. We define the characteristic shock-formation time as the instant when the axion field at the bubble center reaches its first turning point, $\partial_\tau\theta(\tau_c,r=0)=0$, corresponding to $\kappa=M_\phi v_w(\tau_c-\tau_n)$. The resulting values of $\kappa$ are shown in the left panel of Fig.~\ref{fig:geom_plots} as functions of $H_\star/M_\phi$ for different values of $v_w$ and $\theta_i$. We find that $\kappa$ approaches a constant value for $H_\star/M_\phi\ll1$, while it decreases for $H_\star/M_\phi\gg1$. These limiting behaviors admit a simple physical interpretation. The field inside a bubble begins to oscillate once the potential overcomes Hubble friction, i.e.\ when $M_\phi \sim H$. Using the expression for the Hubble parameter in terms of conformal time, $H(\tau)=1/(H_\star\tau^2)$, this condition yields the characteristic shock-formation time $\tau_c\sim(H_\star M_\phi)^{-1/2}$. In the limit $H_\star/M_\phi \gg 1$, we therefore find
\be \label{eq:kappa_large}
    \kappa = M_\phi v_w\,(\tau_c - \tau_n) \sim v_w\left(\frac{M_\phi}{H_\star}
    \right)^{\!1/2} \propto \left(\frac{H_\star}{M_\phi}\right)^{\!-1/2},
\ee
in excellent agreement with the numerical results. In the opposite limit, $H_\star/M_\phi\ll 1$, Hubble friction is negligible from the outset, and the only relevant timescale is $M_\phi^{-1}$, implying $\tau_c\sim M_\phi^{-1}$ and hence $\kappa\sim v_w$.

To estimate the fraction $\varepsilon_{\rm BM}$ of bubbles that contribute to bubble misalignment, we use the distribution of first bubble collision radii derived in Ref.~\cite{Lewicki:2025hxg}
\be \label{eq:p_Rc}
    p(R_c) \propto \int \td t_n \, a(t_n)^3 \Gamma(t_n)\, a(t)\, \bar{F}(t)\,
    \bar{F}'(t)\big|_{t:R(t,t_n)=R_c} \,,
\ee
and integrate over collision radii with $R_c>\kappa/M_\phi$
\be \label{eq:fBM}
    \varepsilon_{\rm BM} = \int_{\kappa/M_\phi}^{\infty} \td R_c \, p(R_c) \,.
\ee
Furthermore, motivated by the analytical result of Ref.~\cite{Lee:2024oaz}, which shows that the axion number density generated through bubble misalignment scales inversely with the characteristic bubble separation, we also compute the mean inverse collision radius,
\be \label{eq:Rcm1}
    \left\langle \frac{1}{R_c} \right\rangle = \frac{1}{\varepsilon_{\rm BM}} \int_{\kappa/M_\phi}^{\infty} \td R_c\, \frac{p(R_c)}{R_c}\,.
\ee
These geometric quantities characterize the bubble misalignment mechanism and depend on the phase transition rate $\beta/H_\star$, as shown in the middle and right panels of Fig.~\ref{fig:geom_plots}. We find that $\varepsilon_{\rm BM}$ decreases with increasing $\beta/H_\star$, reflecting the smaller fraction of bubbles that satisfy the condition $R_c>\kappa/M_\phi$, while $\langle R_c^{-1}\rangle$ increases because the characteristic bubble size becomes smaller. These trends determine the parametric dependence of the bubble misalignment contribution, which we quantify in Sec.~\ref{subsec:fit}.

\begin{figure}
    \centering
    \includegraphics[width=1.0\linewidth]{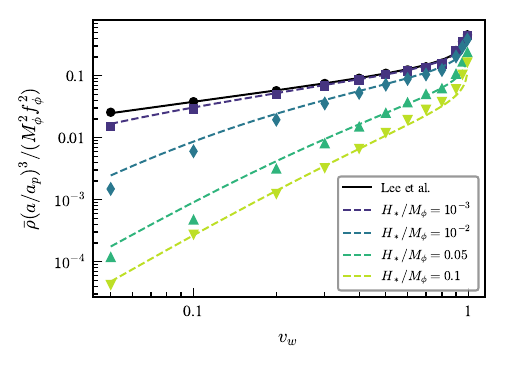}
    \vspace{-4mm}
    \caption{Normalized average axion comoving energy density as a function of the bubble wall velocity $v_w$ for different values of the expansion rate $H_\star/M_\phi$, with $\theta_i=0.1$. The markers show the results of the numerical simulations. The black solid curve corresponds to the result of Ref.~\cite{Lee:2024oaz} obtained in the absence of Hubble expansion, the dashed curves show the Minkowski scaling multiplied by the Hubble dilution factor $(a_p/a_\star)^{-2}$.}
    \label{fig:vw_scaling}
\end{figure}

The final ingredient required to complete our description is the dependence of the axion energy density on the bubble wall velocity. We find that this dependence follows a universal velocity scaling, modified by the effects of cosmic expansion. To quantify it, we perform simulations of a single bubble nucleating at the center of a periodic box and expanding at a constant velocity $v_w$. The average physical energy density is evaluated at the time of complete conversion, when the bubble wall reaches the corners of the box. In the Minkowski limit, $H_\star/M_\phi \to 0$, the energy density follows the power law $\bar{\rho}_\phi \propto (\gamma_w v_w)^p$, where $\gamma_w=(1-v_w^2)^{-1/2}$ and $p\simeq0.6$. The corresponding numerical results, shown by the black solid curve in Fig.~\ref{fig:vw_scaling}, reproduce those of Ref.~\cite{Lee:2024oaz}. In an expanding background, the finite duration of the transition allows Hubble expansion to dilute the radiated shock energy by a factor $(a_p/a_\star)^{-2}$, where $a_\star\equiv a(t_\star)$. This scaling follows because the shock energy is initially stored predominantly in spatial gradients, whose contribution to Eq.~\eqref{eq:averaged-energy-density} scales as $1/\tau^2\propto1/a^2$. Since slower bubble walls produce longer transitions and larger bubbles, the suppression becomes more pronounced as $v_w$ decreases. We find that the velocity dependence is accurately described by the Minkowski scaling multiplied by the Hubble dilution factor $(a_p/a_\star)^{-2}$. This is shown in Fig.~\ref{fig:vw_scaling}, where we plot this combination as a function of the bubble wall velocity for different values of $H_\star/M_\phi$. The agreement between the simulation output and this analytical rescaling is excellent.

\begin{figure*}
    \centering
    \includegraphics[width=0.99\textwidth]{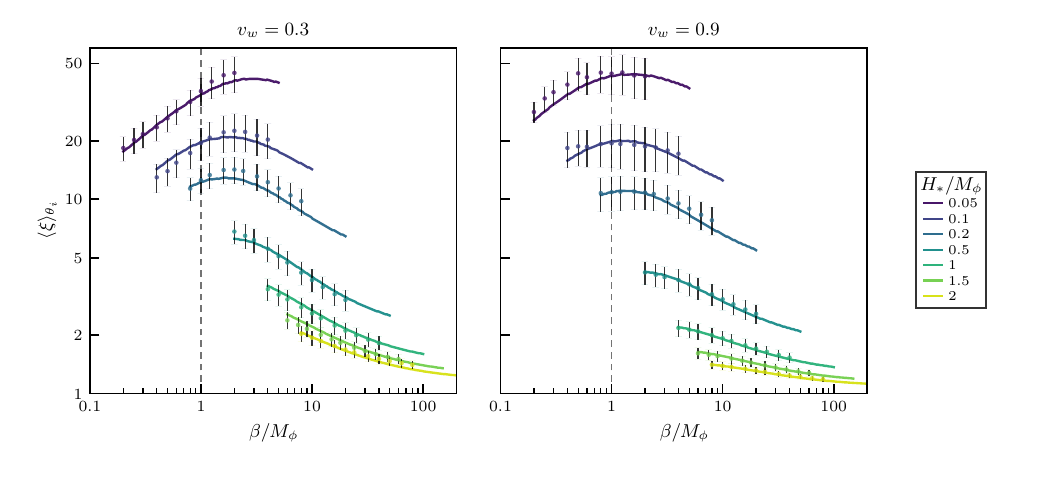}
    \vspace{-7mm}
    \caption{Global fit of the semianalytic model~\eqref{eq:ratio_fit} to the simulation data for bubble wall velocities $v_w=0.3$ (left panel) and $v_w = 0.9$ (right panel). The enhancement factor $\xi$, averaged over initial misalignment angles uniformly distributed in $[0,\pi)$, is shown as a function of $\beta/M_\phi$ for several values of $H_\star/M_\phi$. The solid lines span the range $\beta/H_\star \in [4,100]$. Points with error bars represent simulation results, while solid lines correspond to the best-fit model. The transition from the delayed to bubble misalignment occurs around $\beta/M_\phi \sim 1$ which is shown as a gray dashed line.}
    \label{fig:averaged_fit_results}
\end{figure*}

\subsection{Fit of the enhancement factor}
\label{subsec:fit}

We now combine the results derived in the previous subsections to obtain a semianalytic expression for the axion relic abundance in the presence of a FOPT. We normalize the result to the standard misalignment abundance derived in Sec.~\ref{subsec:standmis} as follows
\be \label{eq:ratio}
    \Omega_\phi^{\rm FOPT} = G\,\xi\,\Omega_\phi^{\rm noFOPT} \,,
\ee
where the total modification is decomposed into two separate factors. The first accounts for the change in the effective number of relativistic degrees of freedom between the conventional oscillation temperature $T_{\rm osc}$ and the percolation temperature $T_p$, and it reads
\be
    G = \frac{g_{*s}(T_{\rm osc})}{g_{*s}(T_p)} \left[ \frac{g_\star(T_p)}{g_\star(T_{\rm osc})} \right]^{3/4},
\ee
The second factor, $\xi=\xi(\theta_i,H_\star/M_\phi,\beta/H_\star,v_w)$, encodes the dynamical impact of the FOPT on the axion abundance, including both delayed and bubble misalignment.

We evaluate $\xi$ directly from numerical simulations by taking the ratio of the late-time comoving axion energy densities, $\bar{\rho}_\phi a^3$, in the phase-transition and no-transition scenarios. The comparison is performed well after the phase transition has completed, when the axion energy density scales as matter in both cases. We perform simulations spanning the parameter ranges $H_\star/M_\phi\in[0.05,2.0]$, $\beta/H_\star\in[4,40]$, $\theta_i\in[0,\pi)$, and $v_w\in[0.3,0.9]$. We use $L_{\rm box}=8\pi/M_\phi$ for low values of $H_\star/M_\phi$ and $L_{\rm box}=2\pi/M_\phi$ otherwise, with $N_{\rm grid}=128$. For each parameter point, we also perform a reference simulation without a phase transition, in which the axion field is initialized uniformly in the true vacuum at $t=0$. The points in Fig.~\ref{fig:averaged_fit_results} show the resulting values of $\xi$, averaged over the initial misalignment angle $\theta_i$.\footnote{A Python implementation of the enhancement factor $\xi$ is publicly available at \url{https://github.com/baltabaygal/axion-fopt}.}

We provide a semianalytic expression for the enhancement factor $\xi$, which allows us both to interpret the numerical results and to extrapolate beyond the simulated parameter region. The model combines the contributions from delayed and bubble misalignment, incorporating geometric information extracted from bubble nucleation and collision statistics. Using the anharmonic correction in Eq.~\eqref{eq:tildefanh} together with the geometric quantities defined in Eqs.~\eqref{eq:fBM} and \eqref{eq:Rcm1}, we model the total enhancement as
\be \label{eq:ratio_fit} 
\xi = \left[\frac{M_\phi}{H_p}\right]^{\frac32} \frac{\widetilde{\mathcal{F}}_{\rm anh}}{\mathcal{F}_{\rm anh}} \!\left[(1 - \varepsilon_{\rm BM})^\lambda + A\, \varepsilon_{\rm BM} \bigg\langle \frac{1}{M_\phi R_c} \bigg\rangle \right], 
\ee
where $A = A_0 (\gamma_w v_w)^p (a_p/a_\star)^{-2}$, with $p \simeq 0.60$, encodes the universal velocity scaling derived in the previous subsection. The first overall factor accounts for the delay in the onset of axion oscillations, while the second accounts for the modification of the anharmonic correction due to the finite onset time of the axion mass. The term in brackets interpolates between the delayed misalignment contribution (dominant for $\varepsilon_{\rm BM}\ll 1$) and the bubble misalignment contribution (dominant for $\varepsilon_{\rm BM}\simeq 1$). For ease of notation, we suppress the arguments of the functions appearing in Eq.~\eqref{eq:ratio_fit}; each function is understood to be evaluated at the appropriate values of the variables defined above.

All parameters entering the semianalytic expression in Eq.~\eqref{eq:ratio_fit} are determined from a global fit to the simulation data, except for the $M_\phi t_p \gg 1$ asymptotic of the anharmonic correction factor, given in Eq.~\eqref{eq:finfty}, which is obtained from the homogeneous evolution. The interpolation parameter $\lambda$ is a constant, $\lambda = 2.27$, and the normalization of the bubble-misalignment term is $A_0 = 0.120$. The broken power-law interpolation of the anharmonic correction~\eqref{eq:tildefanh} is controlled by the parameters $C$ and $r$. We find that $r$ has a systematic $v_w$ dependence that we parametrize as $r(v_w) = r_0 + r_1\, v_w$. The best-fit values are $C = 2.06$, $r_0 = 0.75$ and $r_1 = 0.29$. The dimensionless threshold collision radius is fixed from the 1+1D single-bubble simulations $\kappa_{\rm 1D}$, shown in the left panel of Fig.~\ref{fig:geom_plots}, up to a prefactor: $\kappa = 0.53\, \kappa_{\rm 1D}$. The fit reproduces the simulation data with a log-RMSE of $\sim 10\%$ across the explored parameter space, as seen in Fig.~\ref{fig:averaged_fit_results}. 

\begin{figure}
    \centering
    \includegraphics[width=1.0\linewidth]{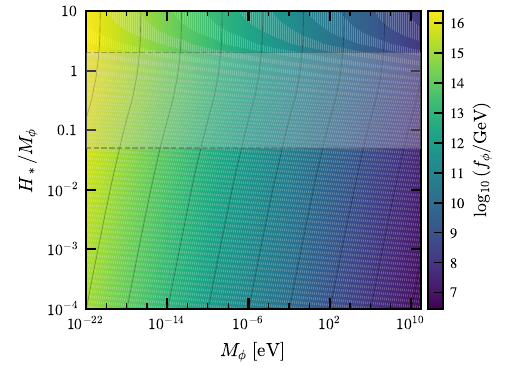}
    \caption{Decay constant $f_\phi$ as a function of axion mass $M_\phi$ and phase 
    transition strength $H_\star/M_\phi$, obtained by averaging over a uniform distribution 
    of initial misalignment angles. The shaded band marks the region $0.05\leq 
    H_\star/M_\phi \leq 2$ where our parametrization has been validated against lattice 
    simulations.}
    \label{fig:mass-H-contour}
\end{figure}

We now turn to the implications for the axion relic density. Figure~\ref{fig:mass-H-contour} illustrates how the enhancement factor $\xi$ modifies the axion decay constant required to reproduce the observed DM abundance. For $H_\star/M_\phi\ll1$, the required decay constant scales as $f_\phi\propto(H_\star/M_\phi)^{1/4}$.

We conclude this section by comparing our semianalytical model and the approximations of Ref.~\cite{Lee:2024oaz} that were obtained for a quadratic potential using simulations of a single bubble in a periodic box without cosmic expansion. Their results can be written as
\bea \label{eq:xi_Lee}
    \xi =
    \begin{cases}
        1\,, & \frac{M_\phi}{3H_p} \leq 1   \\
        \left[\frac{M_\phi}{3 H_p}\right]^{\frac32} , & \frac{M_\phi}{3H_p} > 1 \;{\rm and}\; \frac{\beta}{M_\phi} > 1  \\
        2\,C_v\,v_w^{\,\alpha_v-1} \frac{\beta}{M_\phi} \left[\frac{M_\phi}{3 H_p}\right]^{\frac32} , & \frac{M_\phi}{3H_p} > 1 \;{\rm and}\; \frac{\beta}{M_\phi} \leq 1
    \end{cases} \,,
\eea
where 
\bea
        C_v \simeq
    \left\{
        \begin{array}{ll}
             1.3 \,, & v \lesssim 0.4 \\
             1.8 \,, & v \gtrsim 0.4
        \end{array}
    \right.
    ,
    \quad
    \alpha_v \simeq
    \left\{
        \begin{array}{ll}
             0.4 \,, & v \lesssim 0.4 \\
             0.8 \,, & v \gtrsim 0.4
        \end{array}
    \right.
    .
\eea
Notice that, in the harmonic limit, the field amplitude factors out of the dynamics, so the enhancement depends only on the timing of the transition and the wall velocity. This is consistent with our prediction in Eq.~\eqref{eq:ratio_fit} that reduces to the $\theta_i$-independent form in the small $\theta_i$ limit. 

\begin{figure}
    \centering
    \includegraphics[width=1.0\linewidth]{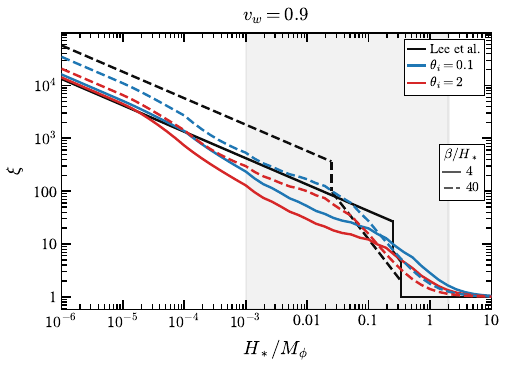}
    \caption{Comparison of the abundance enhancement factor of our semianalytic model given in Eq.~\eqref{eq:ratio_fit} and that of Ref.~\cite{Lee:2024oaz}, given in Eq.~\eqref{eq:xi_Lee}.}
    \label{fig:comparison}
\end{figure}

We compare our results with those of Ref.~\cite{Lee:2024oaz} in Fig.~\ref{fig:comparison}. Qualitatively, the two results are in good agreement. However, even in the harmonic limit we do not reproduce their normalization quantitatively. We attribute this residual offset primarily to calibrating our model using the percolation time $t_p$ measured directly in the simulations rather than the analytic condition $M_\phi=3H$, together with the contribution from inhomogeneous bubble collisions. As expected, the discrepancy becomes more pronounced away from the harmonic limit.

\subsection{Suppressed contribution from cosmic strings}
\label{sec:strings}

The focus of this work is on the misalignment contribution to the axion relic abundance and how a FOPT modifies the standard cosmological prediction. However, if the global symmetry responsible for generating the axion breaks after inflation (the so-called postinflationary scenario), axions are also produced through the decay of the cosmic-string network formed during the spontaneous breaking of the global symmetry. In the standard scenario without a FOPT, axion production from string decay is of $\mathcal{O}(1)$ compared to misalignment~\cite{Vilenkin:1982ks,Kibble:1984hp,Harari:1987ht,Shellard:1987bv,Hagmann:1990mj,Battye:1994au,Kawasaki:2018bzv,Gorghetto:2018myk,Buschmann:2019icd,Hindmarsh:2019csc,Gorghetto:2020qws,Buschmann:2021sdq,Kim:2024wku,Kim:2024dtq,Benabou:2024msj}. We conclude this section by showing that the string-induced contribution is parametrically suppressed in the regime of interest.

It is useful to briefly review what happens in the standard case. Following the PQ symmetry spontaneous breaking, the axion string network rapidly approaches a scaling regime in which the energy density stored in long strings evolves as $\rho_{\rm str}\propto t^{-2}\ln t$ (see, e.g., Ref.~\cite{Sikivie:2006ni}), sustained by the continuous emission of relativistic axions. The onset of the axion mass subsequently leads to the formation of domain walls attached to the strings, after which the string-wall system becomes unstable and rapidly decays. In the standard QCD axion scenario, the mass increases gradually with a power-law dependence on the temperature before reaching its zero-temperature value at the confinement scale. Consequently, the string-wall system decays when $M_\phi\simeq H$, and axion production from strings and from misalignment yield comparable contributions to the final relic abundance.

By contrast, in the framework investigated in this work the axion mass turns on abruptly at $t_\star$, jumping from zero to $M_\phi \gg H_\star$. As a consequence, the relative importance of the two production mechanisms changes dramatically. The energy density carried by axions emitted by the scaling string network at $t_\star$ can be estimated as (see, e.g., Ref.~\cite{Vilenkin:2000jqa})
\be
    \rho_\phi^{\rm str}(t_\star) \simeq c\, \pi \ln\left(\frac{t_\star}{\delta}\right)
    \frac{f_\phi^2}{t_\star^2} \,,
\ee
where $c$ is an order-one dimensionless coefficient and $\delta \sim f_\phi^{-1}$ denotes the string core size. By comparison, the misalignment contribution at $t_\star$ is $\rho_\phi^{\rm mis}(t_\star)\simeq M_\phi^2f_\phi^2$, so that the ratio of the two contributions is
\be
    \frac{\rho_\phi^{\rm str}(t_\star)}{\rho_\phi^{\rm mis}(t_\star)}
    \sim \ln\left(\frac{t_\star}{\delta}\right) \left(\frac{H_\star}{M_\phi}\right)^2 \,.
\ee
The string contribution is therefore suppressed relative to misalignment by a factor of $(H_\star/M_\phi)^2\ll1$, up to the usual logarithmic enhancement from the string tension. Furthermore, the spectrum of axions emitted by the scaling string network satisfies $\partial\rho_\phi^{\rm str}/\partial k\propto1/k$, with an infrared cutoff set by the Hubble scale at which the axion becomes massive~\cite{Gorghetto:2020qws}, corresponding in our case to $k_{\rm min}\sim H_\star$. Since $M_\phi\gg H_\star$, most axions emitted before the phase transition have momenta too small to propagate inside the bubbles, providing an additional suppression of the string contribution. We therefore conclude that axion production from string decay is subdominant throughout the parameter regime considered in this work.

\section{Other Observational Probes}
\label{sec:pheno}

In the previous section, we studied how a FOPT in the early universe modifies the axion relic abundance relative to the standard cosmological scenario. The effect is encapsulated by the enhancement factor defined in Eq.~\eqref{eq:ratio}. Since the relic abundance enters a variety of cosmological and astrophysical observables, the phase transition also leaves characteristic imprints beyond the relic density itself. In this section, we investigate its implications for primordial fluctuations, CMB isocurvature perturbations, and the small-scale cutoff of the matter power spectrum.

The discussion naturally separates into two cases, depending on whether the global symmetry whose spontaneous breaking gives rise to the axion field $\phi$ is broken before or after inflation. In the postinflationary scenario, the phase transition alters the characteristic cutoff scale of the white-noise isocurvature spectrum, thereby affecting the formation and properties of axion miniclusters. In the preinflationary scenario, inflationary fluctuations generate axion isocurvature perturbations whose amplitude is modified by the phase transition dynamics.

\subsection{Axion miniclusters}
\label{subsec:miniclusters}

In the postinflationary scenario, the spontaneous breaking of the global symmetry proceeds through the Kibble mechanism, resulting in an axion field that is approximately homogeneous within each causal horizon while taking uncorrelated values in different horizons. Consequently, the primordial axion fluctuations are characterized by a white-noise power spectrum on scales larger than the horizon at the onset of oscillations.

A useful quantity characterizing the primordial fluctuations is the variance of the initial axion density contrast
\be
    \langle |\delta_\phi|^2 \rangle \equiv \frac{1}{2\pi} \int_{-\pi}^{\pi} \td \theta_i \, \frac{|\rho_\phi(\theta_i)-\bar\rho_\phi|^2}{\bar\rho_\phi^2} \,,
\ee
where $\bar\rho_\phi=(2\pi)^{-1}\int \td\theta_i\,\rho_\phi(\theta_i)$ denotes the average axion energy density. This quantity determines the normalization of the white-noise isocurvature power spectrum, which can be written as~\cite{Fairbairn:2017sil}
\be
    \mathcal{P}_{\rm WN}(k)=\frac{6\pi^2}{D_i(z_{\rm eq})^2}
    k_{\rm cut}^{-3}
    \langle|\delta_\phi|^2\rangle\,
    \Theta(k_{\rm cut}-k)\,.
\ee
Here, $k_{\rm cut}$ denotes the comoving horizon scale at the onset of axion oscillations, while $D_i(z_{\rm eq})$ is the growth factor of isocurvature perturbations evaluated at matter-radiation equality, normalized such that $D_i(z=0)=1$. 

\begin{figure}
    \centering
    \includegraphics[width=1.0\linewidth]{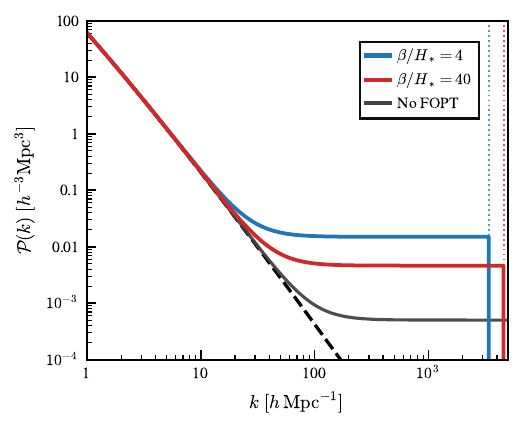}
    \caption{Present-day matter power spectrum for $M_\phi = 10^{-15}$~eV, $H_\star/M_\phi = 0.05$, $v_w = 0.8$. The black dashed curve shows the standard CDM power spectrum $\mathcal{P}_{\rm CDM}(k)$, while the gray solid curve shows the total power spectrum for no FOPT case.}
    \label{fig:Ps-vs-k}
\end{figure}

The total matter power spectrum, including the standard adiabatic cold dark matter (CDM) part $\mathcal{P}_{\rm CDM}(k)$, can be approximated as
\be \label{eq:Ptot}
    \mathcal{P}(k) = \mathcal{P}_{\rm CDM}(k) + \mathcal{P}_{\rm WN}(k)\,.
\ee
At small scales (large $k$),  isocurvature perturbations dominate over the adiabatic component, leading to the formation of axion miniclusters~\cite{Kolb:1993zz}. A phase transition can modify both the characteristic minicluster mass and their abundance by altering the time at which the axion field begins to feel its potential. This effect is captured by the cutoff scale $k_{\rm cut} \approx \min\left[a_{\rm osc} H_{\rm osc}, a_p H_p \right]$, which is shifted to smaller values than in the standard scenario without a phase transition. Consequently, the characteristic scale of the primordial isocurvature fluctuations increases, enhancing the formation of axion miniclusters. This effect is illustrated in Fig.~\ref{fig:Ps-vs-k}.

To give an analytical formula for the scale $k_c$ above which the isocurvature component dominates, we approximate the CDM part in~\eqref{eq:Ptot} as a power-law $\mathcal{P}_{\rm CDM}(k) = \mathcal{P}_{\rm CDM}(k_c) (k/k_c)^n$, where $n$ characterizes the slope of $\mathcal{P}_{\rm CDM}(k)$ at $k \sim k_c$. We find that $n\approx 2.7$ for $k_c = 1 - 1000 h\,{\rm Mpc}^{-1}$ and the transition scale is well approximated by
\be
    \frac{k_{\rm c}}{h\, {\rm Mpc}^{-1}} \simeq \min\!\bigg[ 1.9 \! \left[\tfrac{M_\phi}{10^{-18}\,{\rm eV}}\right]^{3/2n}\!,  3.6 \!\left[\tfrac{H_p}{10^{-18}\,{\rm eV}}\right]^{3/2n} \bigg] .
\ee

At scales $k > k_c$, the matter power spectrum exceeds that of the standard CDM scenario and becomes dominated by the isocurvature fluctuations associated with axion miniclusters. This significantly enhances the high-redshift halo mass function and is therefore constrained by observations of structure formation at high redshift, including Lyman-$\alpha$ forest measurements~\cite{Chabanier:2019eai} and the high-$z$ UV luminosity function~\cite{Irsic:2019iff,Urrutia:2025fvp}. 

\begin{figure}
    \centering
    \includegraphics[width=1.0\linewidth]{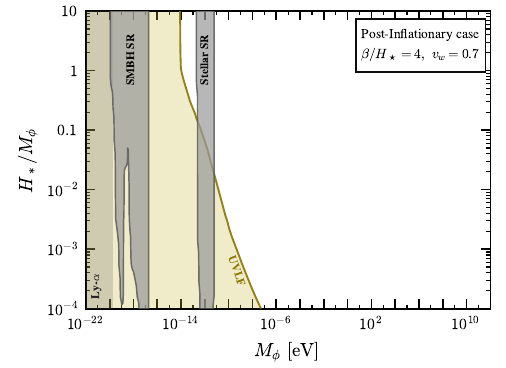}
    \caption{Constraints for the postinflationary case in the $(M_\phi\,, H_\star/M_\phi)$ plane, with $f_\phi$ fixed by requiring that $\phi$ accounts for all DM.}
    \label{fig:mass-vs-h-no-cmb}
\end{figure}

The various observational bounds are summarized in Fig.~\ref{fig:mass-vs-h-no-cmb}, which shows the parameter space in the $(M_\phi,H_\star/M_\phi)$ plane. The axion decay constant $f_\phi$ is fixed so that the axion abundance matches the observed DM abundance, as in Fig.~\ref{fig:mass-H-contour}. The strongest current constraint on the transition scale of the matter power spectrum, $k_c > 37\,h\,{\rm Mpc}^{-1}$, is derived from JWST measurements of the high-redshift UV luminosity function~\cite{Urrutia:2025fvp}. For $H_\star \ll M_\phi$, the UV luminosity function constraint translates into $H_p \gtrsim 3\times10^{-17}\,$eV, corresponding to a phase transition at a temperature $T_p \simeq 0.3\,$MeV. The gray region at the left boundary is excluded by the Lyman-$\alpha$ forest measurements, which constrain $M_\phi > 2\times 10^{-20}\,$eV~\cite{Rogers:2020ltq}. This bound arises from the suppression of small-scale structure below the axion de Broglie wavelength and is therefore independent of $H_\star/M_\phi$. Superradiance constraints from supermassive and stellar-mass black hole spin measurements~\cite{Unal:2020jiy,Witte:2024drg} are also shown and depend on the axion mass $H_\star/M_\phi$ and decay constant $f_\phi$. The apparent dependence on $H_\star$ in Fig.~\ref{fig:mass-vs-h-no-cmb} is entirely indirect, arising through the relic-abundance condition that fixes $f_\phi$.

\subsection{CMB isocurvature}
\label{subsec:pre-inflation}

If the global symmetry is spontaneously broken before or during inflation, the axion behaves as a light spectator field and acquires stochastic fluctuations during inflation~\cite{Graham:2018jyp,Takahashi:2018tdu}, generating a primordial isocurvature component. The amplitude of this component is constrained by cosmic microwave background (CMB) observations to satisfy $P_S(k_{\rm P}) < 8.3\times10^{-11}$~\cite{Planck:2018vyg}, where $k_{\rm P}$ denotes the CMB pivot scale.

A key quantity controlling the stochastic dynamics is the ratio of the axion potential energy to the de Sitter fluctuation scale. We therefore introduce the dimensionless parameter~\cite{Starobinsky:1994bd}
\be \label{eq:alpha-def}
    \alpha \equiv \frac{f_\phi^2 M_\phi^2}{H_I^4} \, .
\ee
In conventional stochastic scenarios~\cite{Jukko:2021hql,Tenkanen:2019aij}, axion fluctuations are strongly suppressed for $\alpha\gg1$. In the framework considered here, however, the axion is massless during inflation and only acquires its mass through the subsequent FOPT. Consequently, the late-time value of $M_\phi$ does not determine the stochastic fluctuations generated during inflation, opening additional parameter space for a stochastic treatment.

\begin{figure}
    \centering
    \includegraphics[width=1.0\linewidth]{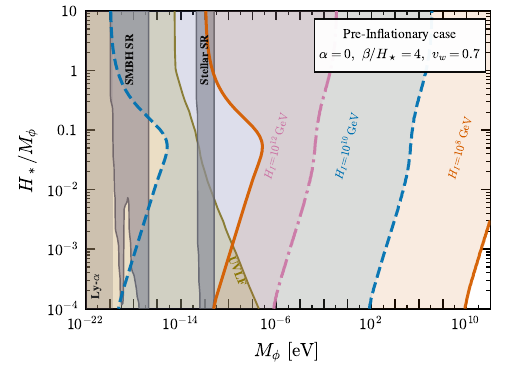}
    \caption{Constraints for the preinflationary stochastic case in the $(M_\phi,H_\star/M_\phi)$ plane, with $f_\phi$ fixed by requiring that $\phi$ accounts for all DM. The UV luminosity function and superradiance constraints are the same as in Fig.~\ref{fig:mass-vs-h-no-cmb}. Additionally, we show the CMB isocurvature constraint $P_S < 8.3\times10^{-11}$ for inflationary scales $H_I = 10^{12}$, $10^{10}$, and $10^{8}$~GeV.}
    \label{fig:mass-vs-h}
\end{figure}

\begin{figure*}
    \centering
    \includegraphics[width=1.0\linewidth]{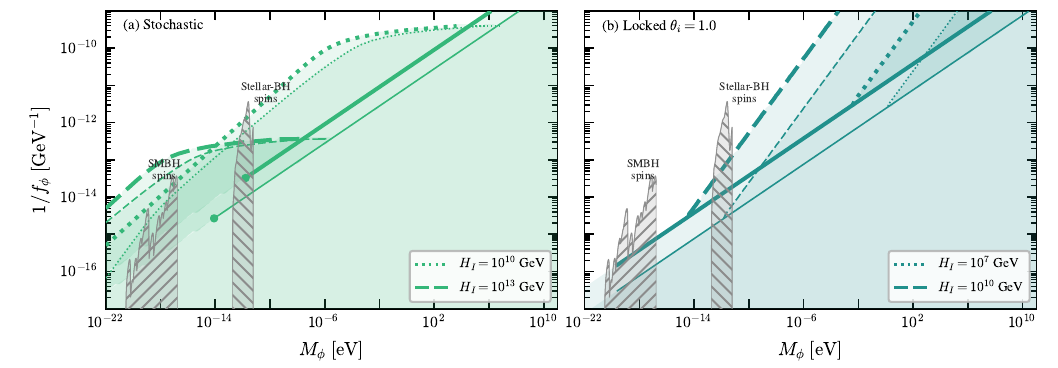}
    \caption{Preinflationary isocurvature constraints for $\beta/H_\star = 4$ and $v_w = 0.7$. In each panel, thin lines show the relic abundance target $\Omega_\phi h^2 \simeq 0.12$ without a phase transition, while thick lines show the result with a phase transition at $H_\star/M_\phi = 0.05$. Hatched regions are excluded by superradiance from supermassive and stellar-mass black holes. The shaded regions are excluded by the CMB isocurvature bound (below the dashed and dotted curves) or by DM overproduction (below the solid curves), for two values of the inflationary Hubble rate $H_I$. Panel (a): stochastic regime, where the axion reaches the equilibrium distribution.  Panel (b): locked regime with fixed initial condition $\theta_i = 1.0$, corresponding to a short period of inflation where the axion distribution can be approximated by a Gaussian of width $\sigma = H_I/(2\pi f_\phi)$. Increasing (decreasing) $\theta_i$ shifts the lines upward (downward).}
    \label{fig:mph-vs-1/fphi_pre_eq}
\end{figure*}

For $\alpha \lesssim 1$ and a sufficiently long period of inflation, the probability distribution of the axion field approaches the stationary solution~\cite{Starobinsky:1994bd}
\be \label{eq:Peq}
    P_{\rm eq}(\theta_i) \propto \exp\!\left[-\frac{8\pi^2}{3 H_I^4} V(\theta_i)\right] \,.
\ee
Following Ref.~\cite{Jukko:2021hql}, we evaluate the isocurvature power spectrum through the spectral decomposition of the Fokker-Planck operator. The eigenfunctions $\psi_n(\theta_i)$ satisfy a Schrödinger-like equation with eigenvalues $\Lambda_n$. In the limit $\alpha=0$, they reduce to $\psi_n(\theta_i)=\cos(n\theta_i)/\sqrt{\pi}$ with dimensionless eigenvalues
\be
    \beta_n \equiv \frac{f_\phi^2 \Lambda_n}{H_I^3} = \frac{n^2}{8\pi^2} \,,
\ee
whereas for $\alpha>0$ the eigenvalue problem is solved numerically. Defining the dimensionless abundance function
\be
    \mathcal{A}(\theta_i) \equiv \theta_i^2\, \mathcal{F}_{\rm anh}(\theta_i)\, \xi(\theta_i) \,,
\ee
which incorporates the effect of the phase transition through the enhancement factor $\xi$ defined in Eq.~\eqref{eq:ratio}, the isocurvature power spectrum at the CMB pivot scale reads
\be \label{eq:Ps-spectral}
    P_S(k_{\rm P}) = \frac{2}{\pi}\sum_{n=1}^{n_{\rm max}} \frac{\mathcal{A}_n^2}{\langle \mathcal{A} 
    \rangle^{\,2}}\, \Gamma\!\left(2-2r\beta_n\right) \sin\!\left(\pi r\beta_n\right) 
    e^{-2r\beta_n N_{\rm P}} \,.
\ee
with spectral coefficients and expectation value
\bea \label{eq:spectral-coeff}
    \mathcal{A}_n &\equiv \int \td \theta_i\, \psi_0(\theta_i)\mathcal{A}(\theta_i)\psi_n(\theta_i) \,, \\
    \langle \mathcal{A} \rangle &\equiv \int \td \theta_i\, P_{\rm eq}(\theta_i) \mathcal{A}(\theta_i) \,.
\eea
and
\be
    r \equiv \left(\frac{H_I}{f_\phi}\right)^{\!2}, \qquad
    N_{\rm P} = 56 + \frac{1}{2} \ln\!\frac{H_I}{8\times10^{13}\,\mathrm{GeV}} \,.
\ee
The resulting constraints in the $(M_\phi,H_\star/M_\phi)$ plane are shown in Fig.~\ref{fig:mass-vs-h}, while the constraints in the $(M_\phi,1/f_\phi)$ plane are presented in the left panel of Fig.~\ref{fig:mph-vs-1/fphi_pre_eq}.

If inflation is too short for the axion field to reach the equilibrium distribution in Eq.~\eqref{eq:Peq}, its probability distribution remains localized around the initial field value $\theta_i$ with a Gaussian width $\sigma = H_I/(2\pi f_\phi)$. In the limit $\sigma \ll 1$, the isocurvature power spectrum reduces to
\be \label{eq:Ps-noneq}
    P_S(k_{\rm P}) \simeq \left[\frac{H_I}{2\pi f_\phi} 
    \frac{\partial\ln\rho_\phi}{\partial\theta_i}\right]^2 \,.
\ee
The resulting CMB exclusion contours are shown in the right panel of Fig.~\ref{fig:mph-vs-1/fphi_pre_eq}.

\section{Conclusions}
\label{sec:conclusion}

The misalignment mechanism remains one of the most compelling scenarios for producing DM. It requires only a light spinless field initially displaced from the minimum of its scalar potential, after which Hubble friction and the onset of coherent oscillations naturally generate a CDM population. Many well-motivated particle candidates possess these properties, with the QCD axion being the prime example because of its connection to the strong $CP$ problem. A key ingredient is the time dependence of the axion mass, which determines the onset of oscillations and hence the final relic abundance. 

In this work, we have investigated how this picture is modified if the axion mass is generated by a FOPT. Initially, the universe resides in the false vacuum, where the axion is massless and frozen by Hubble friction. As true-vacuum bubbles nucleate and expand, the axion acquires a mass and begins evolving according to its scalar potential, giving rise to dynamics that differ significantly from the standard misalignment mechanism.

To study the resulting axion dynamics and relic abundance, we performed lattice simulations in a fully cosmological setting including the effects of cosmic expansion. Across a broad range of phase-transition parameters, we identified two distinct dynamical regimes. For fast transitions, $\beta \gg M_\phi$, the phase transition completes before the axion begins to oscillate. The dominant effect is a delay in the onset of oscillations, determined by the percolation time $t_p$, which enhances the present-day axion abundance relative to the standard misalignment scenario. For slow transitions, $\beta \ll M_\phi$, axion shock waves develop ahead of the expanding bubble walls before the transition completes. The resulting spatial gradients reduce the effective oscillation amplitude inside the bubbles, suppressing the relic abundance relative to the delayed-misalignment regime. This corresponds to the bubble misalignment mechanism first identified in Ref.~\cite{Lee:2024oaz}. Our work extends that analysis by incorporating cosmic expansion, a realistic bubble nucleation history, and anharmonic corrections to the axion potential. While we recover the same qualitative picture, the improved treatment leads to quantitatively different predictions for the relic abundance, as illustrated in Fig.~\ref{fig:vw_scaling}.

In both delayed and bubble misalignment, the net effect of the FOPT is an enhancement of the axion abundance relative to the standard misalignment scenario, shifting the relic-density target toward lower axion masses and smaller decay constants. A key quantity characterizing this effect is the enhancement factor $\xi$, defined in Eq.~\eqref{eq:ratio}, which quantifies the dynamical enhancement over standard misalignment prediction. The remaining factor, $G$, accounts for the change in the effective number of relativistic degrees of freedom due to the delayed onset of oscillations, which occurs at the percolation temperature $T_p$. We derived the semianalytical expression for $\xi$ given in Eq.~\eqref{eq:ratio_fit}, which captures both dynamical regimes through the modified anharmonic correction $\tilde{\mathcal{F}}_{\rm anh}$ and geometric quantities extracted from bubble-collision statistics. This expression reproduces lattice results to within $10\%$ across the parameter space, with only a modest loss of accuracy as $\theta_i \to \pi$.

Besides modifying the relic abundance, the phase transition gives rise to distinctive observational signatures, as discussed in Sec.~\ref{sec:pheno}. These include an enhanced isocurvature power spectrum due to the reduced cutoff scale $k_{\rm cut}$ and a modified population of axion miniclusters. The strongest constraint comes from JWST measurements of the high-redshift UV luminosity function, which require the phase transition to occur above $T_p \simeq 0.3\,\mathrm{MeV}$ for $H_\star \ll M_\phi$. Conversely, the observation of deviations from the standard predictions would provide evidence that the axion mass originated from a FOPT.

The framework here can be extended in several directions. On the phenomenological side, it would be interesting to investigate alternative scalar potentials and assess how different classes of FOPTs modify the axion cosmological evolution and the resulting relic abundance. On the model building side, embedding the mechanism into realistic UV completions could reveal connections with a broader set of cosmological observables, most notably stochastic gravitational-wave backgrounds generated by the same phase transition. This opens the possibility of cross-correlating gravitational-wave measurements with axion observables, such as the relic abundance, isocurvature perturbations, and the properties of axion miniclusters. The most natural extension is to the QCD axion. Besides its compelling theoretical motivation from the strong $CP$ problem, understanding how nonstandard cosmological histories modify the connection between early universe dynamics and low-energy observables is essential for defining the parameter space targeted by future axion searches. This extension is not straightforward: unlike the scenario considered here, the QCD axion acquires a temperature-dependent mass already in the false vacuum through nonperturbative QCD effects. The interplay between this smooth temperature dependence and the abrupt mass generation induced by a FOPT could lead to a richer phenomenology than explored here. Constructing realistic QCD axion models that realize this mechanism and identifying their signatures remains an important direction for future work.

\begin{acknowledgments}
The work of G.B and V.V. was supported by the Estonian Research Council Grants No. TARISTU24-TK3 and No. TARISTU24-TK10, and the Center of Excellence program TK202 of the Estonian Ministry of Education and Research. F.D. and V.V. were supported by Istituto Nazionale di Fisica Nucleare (INFN) through the Theoretical Astroparticle Physics (TAsP) project, and in part by the Italian MUR Departments of Excellence Grants 2023-2027 ``Quantum Frontiers.'' The work of V.V. was also supported by the European Union's Horizon Europe research and innovation program under the Marie Sk\l{}odowska-Curie Grant Agreement No. 101065736. CloudVeneto is acknowledged for the use of computing and storage facilities.
\end{acknowledgments}
\section{Data availability}
The code used to generate and nalyze the data is available from \R{add git to ref}
\appendix

\section{An Illustrative Renormalizable Model}
\label{app:model}

\begin{figure*}
\centering
\includegraphics[width=0.495\linewidth]{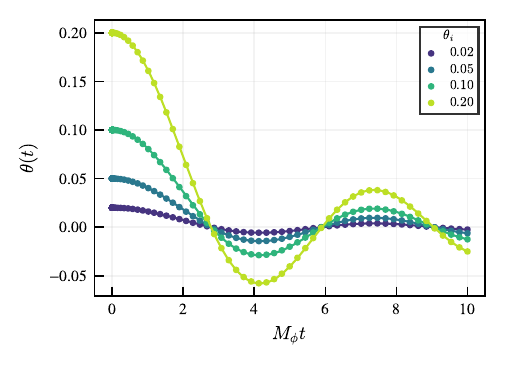}
\includegraphics[width=0.495\linewidth]{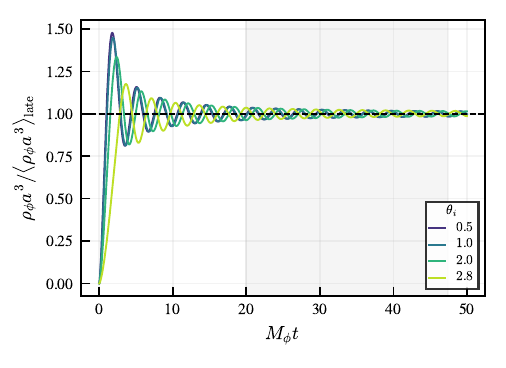}
\caption{Validation of the lattice simulations in the homogeneous limit. Left: linear regime, comparing the analytic solution in Eq.~\eqref{eq:Besssel-solution} (solid lines) with the corresponding lattice results (markers) for several initial axion field values $\theta_i$. Right: nonlinear regime, showing the evolution of the comoving energy density $\rho_\phi a^3$, normalized to its late-time median, for several initial axion field values $\theta_i$.}
\label{fig:homogeneous_validation}
\end{figure*}

The analysis presented in this work is based on the effective scalar potential in Eq.~\eqref{eq:axion_potential}, in which a complex scalar $S$ controls the mass of the axion field $\phi$. This potential is assumed to arise as the low-energy remnant of a renormalizable theory responsible for generating the axion itself. The purpose of this Appendix is to demonstrate that the effective potential employed in the main text admits simple renormalizable UV completions. Since the axion dynamics studied in this work occur well below the masses of the heavy states that have been integrated out, our results are insensitive to the details of the unknown ultraviolet completion.

We assume the existence of two new bosonic fields: a real scalar $S$, which undergoes a FOPT and ultimately controls the axion mass, and a complex scalar $\Phi$, which contains the axion field $\phi$. We divide the full scalar potential into three distinct renormalizable contributions
\be
V_{\rm full}(S,\Phi) = V_0 + V_S(S) + V_\Phi(\Phi) + V_{\rm mix}(S, \Phi) \ .
\label{eq:Vfull}
\ee
The constant term $V_0$ is always chosen to have vanishing potential energy at the global minimum. We impose no restriction on the terms involving only $S$, whereas the sector containing only $\Phi$ is endowed with a global Abelian $U(1)_\Phi$ symmetry under which the scalar transforms as $\Phi \rightarrow \exp[i \alpha] \Phi$. Consequently, the only allowed renormalizable terms are\footnote{We keep the tadpole term for $S$ since we will exploit soon the freedom to perform field redefinitions.}
\begin{align}
\label{eq:VS} V_S(S) = & \, \xi_S S + \frac{\mu_S^2}{2} S^2 + \frac{\kappa_S}{3!} S^3 + \frac{\lambda_S}{4!} S^4 \ , \\
V_\Phi(\Phi) = & \, - \mu_\Phi^2 \Phi^\dag \Phi + \frac{\lambda_\Phi}{4} \left( \Phi^\dag \Phi \right)^2 \ .
\end{align}
If these were the only contributions to the scalar potential, the dynamics of the two fields would be completely decoupled. 

Depending on the choice of parameters, the potential $V_S(S)$ may possess two distinct minima. We redefine the field $S$ by performing an overall shift such that the local minimum is located at $\langle S \rangle = 0$, while the global minimum lies at $\langle S \rangle = v_S \neq 0$. The scalar $S$ can be stabilized at $S = 0$ at early times by thermal effects after reheating and, depending on the relative values of its couplings, may later tunnel through the nucleation of true-vacuum bubbles during the cosmological evolution, thereby generating a FOPT~\cite{Adams:1993zs,Kehayias:2009tn,Gould:2021dzl}. Meanwhile, if $\mu_\Phi^2 > 0$, the minimum of $V_\Phi(\Phi) $ occurs at a nonvanishing vev, implying the spontaneous breaking of the $U(1)_\Phi$ symmetry, with the axion field $\phi$ as the only low-energy remnant. Specifically, denoting the vev by $\langle \Phi \rangle = v_\Phi / \sqrt{2}$, the relation between the complex scalar $\Phi$ and the canonically normalized axion field $\phi$ is
\be
\Phi = \frac{v_\Phi}{\sqrt{2}} \,e^{i \phi/v_\Phi} \ .
\label{eq:PhiExp}
\ee
From this parametrization, it follows that the axion field range is $\phi \in [0, 2 \pi v_\Phi)$. Moreover, in the absence of additional terms in the scalar potential, the axion remains massless at this stage.

Crucially for our work, we also include interactions mixing the two scalar sectors. Imposing renormalizability,  the most general scalar potential takes the form
\be
V_{\rm mix}(S,\Phi) = - \sum_{\scriptscriptstyle n_S \geq 1,\, n_\Phi \geq 1}^{\scriptscriptstyle n_S+n_\Phi \leq 4} 
\frac{\mathcal{C}_{n_S n_\Phi}}{2}\, S^{n_S}\Phi^{n_\Phi} + {\rm H.c.} \ . 
\label{eq:Vmix}
\ee
The sum runs over all integer values satisfying the constraints displayed in the summation symbol. The conditions below enforce the simultaneous presence of both $S$ and $\Phi$, while the condition above follows from renormalizability. Notice how we neglected polynomial contributions with powers of both $\Phi$ and $\Phi^\dag$; once we consider the low-energy theory below the $U(1)_\Phi$ symmetry breaking scale, the combination $\Phi^\dag \Phi$ does not contain the axion field and contributions of this kind just redefine the effective coefficients $\mathcal{C}_{n_S n_\Phi}$ or the coefficients of Eq.~\eqref{eq:VS}.

The operators in Eq.~\eqref{eq:Vmix} explicitly break the global $U(1)_\Phi$ symmetry and therefore generate a nonvanishing axion mass. Using the parametrization in Eq.~\eqref{eq:PhiExp}, the potential can be rewritten in terms of the axion field as
\begin{align}
V(S,\phi) = & \, \sum_{\scriptscriptstyle n_S \geq 1,\, n_\Phi \geq 1}^{\scriptscriptstyle n_S+n_\Phi \leq 4} \Delta V_{n_S n_\Phi}(S,\phi)  \ , \\
\Delta V_{n_S n_\Phi}(S,\phi) = &\, - \frac{\left|\mathcal{C}_{n_S n_\Phi}\right|}{2^{n_\Phi/2}}\, v_\Phi^{n_\Phi} \, S^{n_S} \cos\!\left[\frac{n_\Phi \phi}{v_\Phi} + \delta_{n_S n_\Phi} \right] . 
\end{align}
Here, we have decomposed the potential coefficients into their absolute values and phases $\delta_{n_S n_\Phi}$. 

The scalar potential employed in this work, given explicitly in Eq.~\eqref{eq:axion_potential}, can be reproduced by keeping only terms at constant $n_\Phi$ in the sum above. For example, if we consider real and positive coefficients, corresponding to $\delta_{n_S n_\Phi} = 0$, the matching between the two expressions requires
\begin{align}
f_\phi = & \, \frac{v_\Phi}{n_\Phi} \ , \\
m^2_\phi(S) = & \, \sum_{\scriptscriptstyle n_S \geq 1 }^{\scriptscriptstyle n_S+n_\Phi \leq 4} \frac{\mathcal{C}_{n_S n_\Phi} \, v_\Phi^{n_\Phi} \, S^{n_S}}{2^{n_\Phi/2} f_\phi^2} \ .
\end{align}
Notice how the requirement that the potential is vanishing at the minimum of the theory reproduces not only the cosine part but also the overall constant. In this parametrization, it is also manifest that the axion field range is given by $\phi \in [0, 2\pi n_\Phi f_\phi)$. In the main text of this work, we choose the option $n_\Phi = 1$ such that the field range goes from zero to $2 \pi f_\phi$ and the axion potential has only one physical minimum.

While a complete thermal analysis of this model lies beyond the scope of the present work, we note that both dynamical regimes identified in the main text can be realized within this framework. The axion mass is set by the zero-temperature vev of $S$ and the mixing couplings $\mathcal{C}_{n_S n_\Phi}$, while $H_\star$ and $\beta$ are controlled by the parameters of $V_S(S)$ and their thermal corrections. The mixing couplings contribute only subleadingly to the thermal potential in the natural regime $|\mathcal{C}_{n_S n_\Phi}| \ll \lambda_S$, leaving $\beta/H_\star$ and $M_\phi$ largely independent of $M_\phi$.

\section{Homogeneous Dynamics Validation}
\label{app:validation}

The axion dynamics investigated in this work are intrinsically inhomogeneous. Nevertheless, the homogeneous limit provides a valuable benchmark for the numerical implementation because it admits either exact analytic solutions or robust conservation laws. In this Appendix, we validate the lattice implementation against these benchmarks before applying it to the fully inhomogeneous simulations discussed in the main text.

We initialize the axion field to be spatially homogeneous, such that $\nabla^2\theta=0$ identically, and evolve it in a radiation-dominated background. In this limit, the equation of motion reduces to
\be
\ddot{\theta}+\frac{3}{2t}\dot{\theta}+M_\phi^2\sin\theta=0 \, .
\ee

First, we consider the linear regime, $\theta_i\ll1$, in which the force term in the equation of motion is well approximated by $\sin\theta\simeq\theta$. For regular initial conditions at early times, the equation admits the exact solution 
\be\label{eq:Besssel-solution}
\theta(t)=\theta_i\,2^{1/4}\Gamma\!\left(\frac54\right)(M_\phi t)^{-1/4}J_{1/4}(M_\phi t)\, .
\ee
The left panel of Fig.~\ref{fig:homogeneous_validation} compares this analytic solution (solid lines) with the lattice results (markers) for several small initial amplitudes. The excellent agreement demonstrates that the numerical implementation accurately reproduces both the phase and amplitude of the homogeneous evolution in the linear regime.

As a second test, we consider the nonlinear regime with larger initial field displacements. In this case, an exact analytic solution is no longer available. Nevertheless, once the condition $M_\phi\gg H$ is satisfied, the comoving quantity $\rho_\phi a^3$ must approach a constant, reflecting the conservation of the adiabatic invariant and the transition to matterlike evolution. The right panel of Fig.~\ref{fig:homogeneous_validation} shows $\rho_\phi a^3$, normalized to its late-time median, for several values of $\theta_i$. The expected asymptotic behavior is accurately reproduced in all cases, validating the nonlinear evolution of the homogeneous axion field and the recovery of the correct late-time scaling.

\bibliography{refs}

\end{document}